\documentstyle[12pt,thmsa]{article}

\begin{document}

\author{C. S. Unnikrishnan\thanks{%
Email address: unni@tifr.res.in} \\
{\it Gravitation Group, Tata Institute of Fundamental Research, }\\
{\it Homi Bhabha }Road, {\it Mumbai - 400 005, INDIA},\\
and\\
{\it NAPP Group, Indian Institute of Astrophysics,}\\
{\it Koramangala, Bangalore - 560 034, INDIA}}
\title{Quantum correlations from local amplitudes and the resolution of the
Einstein-Podolsky-Rosen nonlocality puzzle}
\date{Prepared for presentation at the
International Conference on Quantum Optics, ICQO'2000, Minsk, Belarus}
\maketitle

\begin{abstract}
{\normalsize The Einstein-Podolsky-Rosen nonlocality puzzle has been
recognized as one of the most important unresolved issues in the
foundational aspects of quantum mechanics. We show that the problem is more
or less entirely resolved if the quantum correlations are calculated
directly from local quantities which preserve the phase information in the
quantum system. We assume strict locality for the probability amplitudes
instead of local realism for the outcomes, and calculate an amplitude
correlation function.Then the experimentally observed correlation of
outcomes is calculated from the square of the amplitude correlation
function. Locality of amplitudes implies that measurement on one particle
does not collapse the companion particle to a definite state. Apart from
resolving the EPR puzzle, this approach shows that the physical
interpretation of apparently `nonlocal' effects like quantum teleportation
and entanglement swapping are different from what is usually assumed. Bell
type measurements do not change distant states. Yet the correlations are
correctly reproduced, when measured, if complex probability amplitudes are
treated as the basic local quantities. As examples we derive the quantum
correlations of two-particle maximally entangled states and the
three-particle GHZ entangled state.}
\end{abstract}

\section{ Introduction}

Quantum nonlocality as manifested in the EPR correlations has been, without
doubt, the most important unresolved problem in the foundational aspects of
physics. All experiments on quantum correlations and test of Bell's
inequality, and their present interpretations based on the multiparticle
wavefunction in quantum mechanics suggest that there is nonlocality. Yet, we
understand neither the nature of this nonlocality nor the physical mechanism
that establishes the nonlocal correlations. The situation is somewhat akin
to that in the earlier part of the last century when the ether was thought
to be a necessary concept for the propagation of electromagnetic waves, yet
something not apparent or detectable.

Sixty five years ago, Einstein, Podolsky and Rosen (EPR) \cite{epr}
addressed the question whether the wave-function represented a complete
description of reality in quantum mechanics, and argued that it didn't. The
crucial and essential assumption in their argument was strict nonlocality,
in the spirit of special relativity. Also, they had considered and included
the concept of objective reality in the analysis. If the value of an
observable was predictable with certainty without a measurement, the
observable had a physical reality according to EPR. Their assertions lead to
attempts at constructing a hidden variable theory that was more complete.
Bell's analysis of the EPR problem in the early sixties established the
Bell's inequalities obeyed by any local hidden variable theory for the
correlations of entangled particles \cite{bell}. Quantum mechanical
correlations calculated using the entangled wave-function and spin operators
violate these inequalities. Various experiments have established beyond
doubt that there cannot be a viable local realistic hidden variable
description of quantum mechanics \cite{selleri}. Further, these results also
have been interpreted as evidence for nonlocal influences in quantum
measurements involving entangled particles. Since no instruction set carried
by the particles from their source of origin (possibly with the addition of
several local hidden variables) can manage to create the correct
correlations observed in experiments, the only way out seems to be that
measurement of an observable on one of the particles in an entangled pair
seems to convey the result of this measurement instantaneously to the other
particle resulting in the correct behaviour of the other particle during a
measurement on the second particle. In the quantum mechanical terminology,
the measurement of an observable on one of the particles collapses the
entire wave-function instantaneously and nonlocally and the second particle
acquires a definite value for the same observable, consistent with the
relevant conservation law. The no signalling theorems in this context
prohibit any faster than light signalling using this feature, and therefore
signal locality is not violated. But the stronger requirement of Einstein
locality is violated. We seem to be stuck with the puzzling nonlocality
which is probably the deepest mystery in the behaviour of entangled systems.

Accepting the concept of nonlocality without being able to understand its
nature is already a disturbing feature. There is also serious conflict with
the spirit of relativity. If one measurement precede the other in one frame,
one can always find a moving frame in which the converse it true, the second
measurement preceding the first \cite{penrose}. Therefore, one cannot
attribute cause and effect relationships for measurements causing nonlocal
collapse.

It turns out that the long-standing problem of EPR nonlocality is resolved
by a simple quantum step that is physically well motivated \cite{unni1}. The
crucial new idea is to incorporate the fact that quantum systems have `wave
aspects' in their behaviour and all calculations should take into account
the phase relationships (coherence) that might be there in the multiparticle
system. If locality is assumed at the level of probability amplitudes, and
if the correlation is calculated directly from these amplitudes the correct
quantum correlations emerge. This means that the EPR definition of objective
reality was too restrictive. There is objective reality, but that is at the
level of quantum phases and not at the level of eigenvalues. The quantum
correlation is encoded in the relative phase appropriate for the problem. In
the local hidden variable theories the correlations are calculated from
eigenvalues and this procedure {\em does not preserve the phase information}%
. The situation has some analogy to the description of interference in
quantum mechanics. Any attempt to reproduce the interference pattern using
locality and the information on `which-path' will fail since the phase
information is lost or modified in such an attempt.

\section{The solution of the EPR nonlocality puzzle}

Consider the breaking up of a correlated state as in the standard Bohm
version of the EPR problem \cite{bohm}. The two-particle state is described
by the wave function 
\begin{equation}
\Psi _{S}=\frac{1}{\sqrt{2}}\{\left| 1,-1\right\rangle -\left|
-1,1\right\rangle \}
\end{equation}

where the state $\left| 1,-1\right\rangle $ is short form for $\left|
1\right\rangle _{1}\left| -1\right\rangle _{2\newline
}$, and represents an eigenvalue of $+1$ for the first particle and $-1$ for
the second particle{\em \ if measured} in any particular direction. $\Psi
_{S}$ is inherently nonlocal, describing both particles together, even when
they are far apart in space-like separated regions.

Two observers make measurements on these particles individually at space
like separated regions with time stamps such that these results can be
correlated later through a classical channel. We assume that strict locality
is valid at the level of probability amplitudes \cite{unni1}. A measurement
changes probability amplitudes only locally. {\em Measurements performed in
one region do not change the magnitude or phase of the complex amplitude for
the companion particle in a space-like separated region. }The local setting
of the polarizers, analyzers, Stern-Gerlach analyzers etc. (collectively
denoted as analyzer) is represented by ${\bf a}$ and ${\bf b}$ for the two
distant apparatus. These could be the directions of the analyzers, for
example. Since we need to deal with correlated particles which may have a
definite phase relationship at source (when the particles are produced
together, for example) we introduce internal variables associated with each
particle. We denote these variable as $\phi _{1}$ and $\phi _{2}$. Their
values are unaltered once the particles are separated. Measurement on one
particle does not change the value of this internal variable for the other
particle. The assumption of locality is that the amplitudes (as opposed to
eigenvalues) are functions of only these local variables.

We now state the assumptions (1 and 2) and some related comments (3, 4,and
5) :

\begin{enumerate}
\item  The local amplitude for the first particle $C_{1}$ that decide the
passage of the particle through an analyzer depends only on the local
variables ${\bf a}$ and $\phi _{1}.$ Similarly $C_{2}$ depends only on $b$
and $\phi _{2}$. If we denote the passage as $+$ and the alternate outcome
as $-,$ then the statement of locality is for the relevant amplitudes is 
\begin{equation}
C_{1\pm }=C_{1\pm }({\bf a},\phi _{1}),\quad C_{2\pm }=C_{2\pm }({\bf b}%
,\phi _{2})
\end{equation}

\item  The correlations of the particles are encoded in the difference of
the internal variables $\phi _{1}$ and $\phi _{2}.$ If the particles have
perfect correlations at source then all the pairs in the ensemble have the
same value for the difference $\left| \phi _{1}-\phi _{2}\right| =\phi _{0}.$

\item  We do not make any assumption on determinism. Given the initial
values of the internal variables $\phi _{1}$ and $\phi _{2}$, we do not
attempt to make any prediction of the eigenvalues that would be measured in
each run of the experiment.

\item  We do not assume any hidden variable in the problem. The variables $%
\phi _{1}$ and $\phi _{2}$ are associated with the particles and they could
be considered as hidden variables in a formal sense, though these values are
not measurable since only relative phases are measurable.

\item  We will also state the locality at the level of the eigenvalues,
though we do not use this in the calculation. For observables $A$ and $B$,
\end{enumerate}

\begin{equation}
A({\bf a,}\phi _{1})=\pm 1,\quad B({\bf b,}\phi _{2})=\pm 1
\end{equation}

This is the same locality assumption as in local realistic theories \cite
{bell}. But, this has a meaning different from its meaning in standard local
realistic theories. Here, this means that the outcomes, {\em when measured},
depend only on the local setting and the local internal variable. There is
no objective reality to $A$ and $B$ before a measurement. There is objective
reality to $\phi _{1}$ and $\phi _{2},$ but there is no way to observe these
absolute phases.

Note that $\phi $ is not a dynamical phase evolving as the particle
propagates. It is an internal variable whose difference (possibly zero)
remains constant for the particles of the correlated pair. The value of $%
\phi $ can vary from particle to particle, but the relative phase $\phi _{0}$
between the two particles in all correlated pairs is constant. Consider $%
\phi $ as a reference for the particles to determine the angle of a
polarizer or analyzer encountered on their way, {\em locally.}

The first particle encounters analyzer1 kept at an angle $\theta _{1}$ with
respect to some global direction. We denote this angle of the analyzer with
reference to $\phi $ as $\theta .$ Similarly, the second particle which has
the internal phase angle $\phi +\phi _{0},$ where $\phi _{0}$ is a constant,
encounters the second analyzer oriented at angle $\theta _{2}$ at another
space-like separated point. Let the orientation of this analyzer with
respect to the internal phase angle of the second particle is $\theta
^{\prime }.$ We have $\theta -\theta ^{\prime }=\theta _{1}-\theta _{2}+\phi
_{0}.$

An experiment in which each particle is analyzed by orienting the analyzers
at various angles $\theta _{1}$ and $\theta _{2}$ is considered next. At
each location the result is two-valued denoted by ($+1$) for transmission
and ($-1$) for absorption of each particle, for any angle of orientation.
The classical correlation function, which is also the experimenter's
correlation function, $P({\bf a},{\bf b})=\frac{1}{N}\sum (A_{i}B_{i})$
satisfies $-1\leq P({\bf a},{\bf b})\leq 1.$ Here $({\bf a},{\bf b})$
denotes the two directions along which the analyzers are oriented and $A_{i}$
and $B_{i}$ are the two valued results. We note that $P({\bf a},{\bf b})$
denotes the average of the quantity ({\em number of} {\em detections in
coincidence }$-${\em \ number of detections in anticoincidence}), where
`coincidence' denotes both particles showing same value for the measurement
and `anticoincidence' denotes those with opposite values. The defect in the
local realistic theories is that they try to calculate this correlation
essentially by averaging over the products of eigenvalues. Obviously the
phase information is thrown away in this procedure and there is no way,
conceptually, such an attempt would have reproduced quantum mechanical
results. We calculate the experimenter's correlations starting from local
amplitudes.

Now we state the expressions for the amplitudes and the amplitude
correlation function. This is new physical input \cite{unni1}.

\begin{enumerate}
\item  The local amplitudes for transmission associated with the first
particle is $C_{1+}=\frac{1}{\sqrt{2}}\exp (i\theta s)$ for measurements at
analyzer1, and for the second particle it is $C_{2+}=\frac{1}{\sqrt{2}}\exp
(i\theta ^{\prime }s)$ at analyzer2. The amplitudes for the orthogonal
outcome are $C_{1-}$ and $C_{2-}$ and these are rotated by $\pi /2$ in the
complex plane from the amplitudes for transmission. The square of the
amplitudes give the corresponding probabilities ($C_{1+}C_{1+}^{*}$ gives
the probability for transmission, for example). $s$ is the spin of the
particle (1 for photons and $\frac{1}{2}$ for the spin-$\frac{1}{2}$ singlet
state- see below)

\item  The amplitude correlation function is a normalized inner product of
the amplitudes. This is of the form 
\begin{equation}
U({\bf a},{\bf b})=~Real(NC_{i}C_{j}^{*})
\end{equation}
where $N$ is the normalization constant. The square of the amplitude
correlation function gives the joint probabilities for events of the form $%
(++),$ $(--),$ $(+-),$ and $(-+).$ All probabilities are guaranteed to be
positive definite in our formalism since the amplitude correlation function
is real.
\end{enumerate}

{\em The crucial difference from local realistic theories is that the
correlation is calculated from quantities which preserve the relative
phases. }

Let us consider the maximally entangled singlet system described by Eq. 1,
the most widely discussed example in the context of nonlocality. We
prescribe the local amplitudes as $C_{1+}=\frac{1}{\sqrt{2}}\exp \{is(\theta
_{1}-\phi _{1})\}$ for the first particle at the first polarizer and $%
C_{2+}= $ $\frac{1}{\sqrt{2}}\exp \{is(\theta _{2}-\phi _{2})\}$ for the
second particle at the second polarizer. There are corresponding amplitudes, 
$C_{1-} $ and $C_{2-}$ for the events denoted by $-,$ and they differ only
in the phase for the maximally entangled state.

The explicit dependence of the amplitude on the spin of the particle is
motivated by the fact that we are dealing with systems with phases and the
phase associated with the spin rotations (a geometric phase) is a necessary
input in this description \cite{unni2}. The correlation at source is encoded
in $\phi _{0}$. The locality assumption is strictly enforced since the two
amplitudes depend only on local variables and on an internal variable
generated at the source and then individually carried by the particles
without any subsequent interaction of any sort. The individual measurements
at each end separately will now give the correct result for transmission for
any angle of orientation. These probabilities are 
\begin{equation}
C_{1}C_{1}^{*}=C_{2}C_{2}^{*}=\frac{1}{2}
\end{equation}

Events of both types ($++$) and ($--$) contribute to a ``coincidence''. The
correlation function for an outcome of either $(++)$ or $(--)$ of two
maximally entangled particles is 
\begin{equation}
U(\theta _{1},\theta _{2},\phi _{o})=2\ {Re}(C_{1}C_{2}^{*})=\cos \{s(\theta
_{1}-\theta _{2})+s\phi _{o}\}.
\end{equation}

It is normalized such that its square will give the conditional joint
probabilities of the type `outcome $+$ for the second particle, given that
the outcome for the first particle is $+,$ etc. All references to the
individual values of the internal variable $\phi $ has dropped out.

We now derive the relation between this correlation function and the
experimenter's correlation function $P({\bf a},{\bf b})=\frac{1}{N}\sum
(A_{i}B_{i})$. Since $U_{++}^{2}=U_{--}^{2}$ for the maximally entangled
state, $U^{2}(\theta _{1},\theta _{2},\phi _{o})$ is the probability for a
coincidence detection ($++$ or $--$), and $(1-U^{2}(\theta _{1},\theta
_{2},\phi _{o}))$ is the probability for an anticoincidence (events of the
type $+-$ and $-+$). Since the average of the quantity (number of
coincidences $-$ number of anticoincidences) = 
\begin{equation}
U^{2}(\theta _{1},\theta _{2},\phi _{o})-(1-U^{2}(\theta _{1},\theta
_{2},\phi _{o}))=2U^{2}(\theta _{1},\theta _{2},\phi _{o})-1,
\end{equation}
the correspondence between $P({\bf a},{\bf b})$ and $U(\theta _{1},\theta
_{2},\phi _{o})$ is given by the expression, 
\begin{eqnarray}
P({\bf a},{\bf b}) &=&2U^{2}(\theta _{1},\theta _{2},\phi _{o})-1  \nonumber
\\
&=&2\cos ^{2}\{s(\theta _{1}-\theta _{2})+s\phi _{o}\}-1
\end{eqnarray}

This completes the back-bone of our formalism and we are ready to discuss
some specific examples.

\section{Spin-$\frac{1}{2}$ particles and Photons}

Consider the singlet state breaking up into two spin-$\frac{1}{2}$ particles
propagating in opposite directions to spatially separated regions. Since
orthogonality of the two particles in any basis implies a relative angle of $%
\pi $ for spinors, we set $\phi _{o}=\pi $ . Then the correlation function
and $P({\bf a},{\bf b})$ calculated from this function are

\begin{eqnarray}
U(\theta _{1},\theta _{2},\phi _{o}) &=&\cos \{s(\theta _{1}-\theta
_{2})+s\phi _{o}\}  \nonumber \\
&=&\cos \{\frac{1}{2}(\theta _{1}-\theta _{2})+\pi /2\}  \nonumber \\
&=&-\sin \frac{1}{2}(\theta _{1}-\theta _{2})
\end{eqnarray}

\begin{eqnarray}
P({\bf a},{\bf b}) &=&2\sin ^{2}(\frac{1}{2}(\theta _{1}-\theta _{2}))-1 
\nonumber \\
&=&-\cos (\theta _{1}-\theta _{2})=-{\bf a}\cdot {\bf b}
\end{eqnarray}

This is identical to the quantum mechanical predictions obtained from the
singlet entangled state and Pauli spin operators. We have reproduced the
correct correlation function using local amplitudes.

For the case of photons entangled in orthogonal polarization states we get,
by setting $s=1$ and $\phi _{o}=\pi /2$ to represent orthogonal
polarization, 
\begin{eqnarray}
U(\theta _{1},\theta _{2},\phi _{o}) &=&\cos \{(\theta _{1}-\theta _{2})+\pi
/2\}  \nonumber \\
&=&-\sin (\theta _{1}-\theta _{2})
\end{eqnarray}

\begin{equation}
P({\bf a},{\bf b})=2\sin ^{2}(\theta _{1}-\theta _{2})-1=-\cos (2((\theta
_{1}-\theta _{2}))
\end{equation}
which is the correct quantum mechanical correlation.

The same analysis works for particles entangled in other sets of variables
like momentum and coordinate, and energy and time. These cases of two
particle entanglement can be mapped on to the spin-$\frac{1}{2}$ singlet
problem with two-valued outcomes. Starting from the local amplitudes $C_{1}=%
\frac{1}{\sqrt{2}}\exp (i\alpha k(x_{1}-x_{o})/2),$ and $C_{2}=\frac{1}{%
\sqrt{2}}\exp (i\alpha k(x_{2}-x_{o})/2)$ we can derive the probability for
coincidence detection as 
\begin{equation}
P(x_{1},x_{2})=\cos ^{2}(\alpha k(x_{1},x_{2})/2)=\frac{1}{2}(1+\cos k\alpha
(x_{1}-x_{2}))
\end{equation}
This is the two photon correlation pattern with 100\% visibility, {\em %
obtained without nonlocality}. $x_{1}$ and $x_{2}$ are the coordinates of
the two detectors separated by a space-like interval. $k$ is the wave vector
and $\alpha $ is a scaling factor for the angle subtended by the two slits
at the detectors, source etc. The factor $2$ dividing the angular variable
comes from the mapping with the spin-$\frac{1}{2}$ problem.

\section{Three-particle GHZ correlations}

The three particle G-H-Z state \cite{ghz} is defined as 
\begin{equation}
\left| \Psi _{GHZ}\right\rangle =\frac{1}{\sqrt{2}}(\left|
1,1,1\right\rangle -\left| -1,-1,-1\right\rangle )
\end{equation}

where the eigenvalues in the kets are with respect to the $z$-axis basis.

The prediction from quantum mechanics for the measurement represented by the
operator $\sigma _{x}^{1}\otimes \sigma _{x}^{2}\otimes \sigma _{x}^{3}$ is
given by 
\begin{equation}
\sigma _{x}^{1}\otimes \sigma _{x}^{2}\otimes \sigma _{x}^{3}\left| \Psi
_{GHZ}\right\rangle =-\left| \Psi _{GHZ}\right\rangle
\end{equation}

Equivalently the joint probabilities for various outcomes in the $x$
direction are 
\begin{eqnarray}
P(+,+,+) &=&P(-,-,+)=P(+,-,-)=P(-,+,-)=0 \\
P(-,-,-) &=&P(+,+,-)=P(+,-,+)=P(-,+,+)=1
\end{eqnarray}
Local realistic theories predict that the product of the outcomes in the $x$
direction for the three particles should be $+1,$ i.e.e, 
\[
P(+,+,+)=P(-,-,+)=P(+,-,-)=P(-,+,-)=1 
\]
This contradicts Eqs. 15-17 and highlights the conflict between a local
realistic theory and quantum mechanics.

The solution using local amplitudes is simple and physically revealing \cite
{unni3}. We define the local amplitudes for the outcomes $+$ and $-$ at the
analyzer (with respect to the $x$ basis) for the first particle as $C_{1+}=%
\frac{1}{\sqrt{2}}\exp (i\theta _{1})$, and $C_{1-}=\frac{1}{\sqrt{2}}\exp
(i(\theta _{1}+\pi /2)).$ The amplitude $C_{1-}$ contains the added angle $%
\pi /2$ because this amplitude is orthogonal to $C_{1+}.$ Similarly, we have 
$C_{2+}=\frac{1}{\sqrt{2}}\exp (i\theta _{2})$, and $C_{2-}=\frac{1}{\sqrt{2}%
}\exp (i(\theta _{2}+\pi /2))$ for the second particle and $C_{3+}=\frac{1}{%
\sqrt{2}}\exp (i\theta _{3})$, and $C_{3-}=\frac{1}{\sqrt{2}}\exp (i(\theta
_{3}+\pi /2))$ for the third particle.

Correlation function is obtained from $N$Real($C_{1}C_{2}^{*}C_{3}^{*}),$
where $N$ is a normalization constant, and its square is the relevant joint
probability. (There is no unique definition of the amplitude correlation
function. The final results are independent of the particular definition we
use). Since we want $N$Re$(C_{1-}C_{2-}^{*}C_{3-}^{*})=\pm 1,$ we choose $%
C_{1-}C_{2-}^{*}C_{3-}^{*}$ to be pure real$.$ This gives 
\[
\frac{N}{2\sqrt{2}}Real(\exp i(\theta _{1}-\theta _{2}-\theta _{3}-\pi
/2))=\pm 1 
\]
\[
\theta _{1}-\theta _{2}-\theta _{3}-\pi /2=0{\rm {\ or}\pm \pi } 
\]
We can choose the relevant relative phases to satisfy this condition. Then
we get 
\[
P(-,-,-)=1 
\]

Rest of the joint probabilities given in Eq. 6 automatically follow, since
flipping sign once rotates the complex number $C_{1-}C_{2-}^{*}C_{3-}^{*}$
through $\pi /2.$ The square of $N$Real($C_{1}C_{2}^{*}C_{3}^{*})$ is then $%
1 $ for an odd number of $(-)$ outcomes and $0$ for even number of $(-)$
outcomes.

Similar construction also applies to four- particle maximally entangled
state \cite{sam} and general multiparticle maximally entangled states.

\section{Concluding remarks}

We have also constructed local amplitudes for the Hardy experiment \cite
{hardy} in which quantum mechanics predicts three particular zero joint
probabilities are one nonzero joint probability (the other possible joint
probabilities in the problem can be nonzero and are not relevant for the
demonstration of nonlocality). Local complex amplitudes that reproduce the
four relevant joint probabilities can be constructed easily. It is
impossible to achieve this if local realism at the level of eigenvalues are
assumed.

We note that there is a simple way to physically understand the fact that
the quantum correlations are typically larger than the corresponding
classical correlations. The overlap (inner product) between a normalized
random vector with $N$ elements and any basis vector is $1/N$ for a
classical vector, and $1/\sqrt{N}$ for a quantum vector (amplitude).
Therefore quantum correlations are typically stronger than classical
correlations. In fact, it is this same physical fact that forms the basis of
quantum search algorithms \cite{grover}, where the initial overlap between a
random vector and the desired basis vector is $\sqrt{N}$ times larger in the
quantum case, making the search faster by $\sqrt{N}.$

The following table summarizes the locality and reality properties in
various approaches to quantum correlations:

\medskip

\begin{tabular}{|p{0.9in}|p{0.9in}|p{0.6in}|p{0.6in}|c|c|}
\hline\hline
\multicolumn{1}{||p{0.9in}|}{Theory/ Formalism} & Basic quantity & Locality
& Reality & Determinism & \multicolumn{1}{|c||}{Predictions} \\ \hline\hline
Quantum mechanics & Multiparticle wavefunction & NO & NO & 
\multicolumn{1}{|l|}{NO} & \multicolumn{1}{|l|}{Correct} \\ \hline
Local Realistic theories with hidden variables & Eigen values & YES & YES & 
\multicolumn{1}{|l|}{YES} & \multicolumn{1}{|l|}{Incorrect} \\ \hline
Present formalism & Amplitudes & YES & Yes (for phase) & 
\multicolumn{1}{|l|}{NO} & \multicolumn{1}{|l|}{Correct} \\ \hline
\end{tabular}

\medskip

Quantum entanglement swapping \cite{swap} is understood within this frame
work by noting that Bell state measurements choose subensembles of particle
pairs that show a particular joint outcome. Particles entangled
independently with the pair of particles that are subjected to the Bell
state measurement will show a joint outcome consistent with swapped
entanglement due to the correlation encoded in the internal variable. {\it %
But the Bell state measurement does not collapse the distant particle into a
definite state}. Yet all correlations are correctly reproduced. This has
important implication to the interpretation of quantum teleportation. The
present nonlocal interpretation of quantum teleportation is not correct.

In summary, the long standing puzzle of nonlocality in the EPR correlations
is resolved. There is no nonlocal influence between correlated particles
separated into space-like regions. The solution has new physical and
philosophical implications regarding the nature of reality, measurement and
state reduction in quantum systems.


\begin{thebibliography}{99}
\bibitem{unni2}  C. S. Unnikrishnan, to be published; The geometric phase
associated with spin rotations can be obtained directly from a
generalization of the usual dynamical phase $\int p_{i}dx^{i},$ to include
the spin as one of the momenta $p_{i}$ and the angular coordinate as the
corresponding $dx^{i}$.
\end{thebibliography}
\end{document}